\newcommand{\ncm}{\newcommand}
\ncm{\cstar}{$C^{*}$-algebra} 
\ncm{\cstars}{$C^{*}$-algebras} 
\ncm{\Ga}{\mbox{$\Gamma$}} 
\ncm{\ra}{\rightarrow} 
\ncm{\st}[2]{\stackrel{#1}{#2}} 
\ncm{\ral}{\longrightarrow} 
\ncm{\al}{\mbox{$\alpha $}} 
\ncm{\gam}{\mbox{$\gamma $}} 
\ncm{\vp}{\varphi} 
\ncm{\vep}{\varepsilon} 
\ncm{\vt}{\vartheta} 

\ncm{\Ad}{\mbox{\rm Ad}} 

\ncm{\mod}{\mbox{\rm mod}} 
\ncm{\ol}{\overline} 

\ncm{\OA}[1]{{\cal O}_{#1}} 
\ncm{\OPA}[1] {\mbox{$\bar{\cal O}_{#1}$}} 
\ncm{\FA}[1]{{\cal F}_{#1}} 

\ncm{\SFA}[1] {\mbox{$\bar{\cal F}_{#1}$}} 
\ncm{\EA}[1]{{\cal E}_{#1}} 
\ncm{\KO}[2]{K_{#1} (\OA{#2})} 
\ncm{\KE}[2]{K_{#1} (\EA{#2})} 

\ncm{\ld}{\lneqq} 
\ncm{\lb}{\label} 
\ncm{\qref}[1]{(\ref{#1})}

\ncm{\qed}{\hfill $\square$ \smallskip} 

\documentclass[10pt]{article}
\usepackage{latexsym,amssymb}
\usepackage{graphicx}
\usepackage{subcaption}

\newtheorem{theo}{Theorem}[section]

\newtheorem{notation}[theo]{Notation}

\newtheorem{remark}[theo]{Remark}
\newtheorem{example}[theo]{Example}
\newtheorem{definition}[theo]{Definition}
\newtheorem{problem}[theo]{Problem}

\newtheorem{zadatak}[theo]{Zadatak}

\newenvironment{jbas}{\begin{equation}}{\end{equation}}

\newenvironment{jba}{\begin{displaymath}}{\end{displaymath}}

\ncm{\R}{\mathbb R} 
\ncm{\E}{\mathbb E}

\ncm{\Z}{\mathbb Z} 
\ncm{\T}{\mathbb T} 
\ncm{\TT}{\T$^{2}$} 

\ncm{\N}{\mathbb N} 
\ncm{\C}{\mathbb C} 

\ncm{\K}{\mathbb K}

\usepackage{hyperref}
\usepackage{placeins}

\usepackage{booktabs}
\usepackage{caption}
\usepackage{colortbl}
\usepackage{xcolor}
\usepackage{multirow}

\definecolor{LightCyan}{rgb}{0.88,1,1}
\definecolor{Gray}{gray}{0.9}

\begin{document}

\begin{center} {\Large  A Clique-Based Method for Improving Motif Scanning Accuracy, v3.01\\} 
\end{center} \bigskip \begin{center} {Braslav Rabar, Keti Ni\v{z}eti\'{c}  and Pavle Goldstein  \\ University of Zagreb, Faculty of Science, Mathematics Department } 
\end{center}
\bigskip
\section*{Abstract}
\subsection*{Background} Motif scanning is a very common method in bioinformatics. Its 
objective is to detect motifs of sufficient similarity to the query, which is then used to 
determine familiy membership, or structural or functional features or assignments. 
Considering a variety of uses, accuracy of motif scanning procedures is of great 
importance. 
\subsection*{Results} We present a new approach for improving motif scanning accuracy, based on 
analysis of in-between similarity. Given a set of motifs obtained from a scanning process, 
we construct an associated weighted graph. We also compute the expected weight of an 
edge in such a graph. It turns out that restricting results to the maximal clique in 
the graph, computed with respect to the expected weight, greatly increases precision, hence improves accuracy of the scan. 
We tested the method on an ungapped motif-characterized protein family from five plant proteomes. The method 
was applied to three iterative motif scanners - PSI-BLAST, JackHMMer and IGLOSS - with 
very good results. 
\subsection*{Conclusions} We presented a method for improving protein motif scanning accuracy, and 
have successfully applied it in several situations. The method has wider implications, 
for general pattern recognition and feature extraction strategies, as long as one can 
determine the expected similarity between objects under consideration.

\section{Background}

Motif scanning - or local similarity search - is a very important part of sequence analysis. It can be used for various purposes - protein family assignment (\cite{Finn2007}), secondary structure prediction (\cite{PSSP}) and similar. Motif scanning methods are typically based on a local alignment algorithm - Smith-Waterman or Viterbi algorithm - with various modifications added, such as approximations and variations in the scoring function or model building. 

Motif scanning procedures normally take an instance of a motif - or even a profile - as an input - that is a {\em query} - and search for similar patterns in a set of sequences. The output consists of a set of sufficiently similar matches, and these form a set of positives, or - what we call it here - the {\em response}. 

In this paper, we are concerned with accuracy of motif scanning procedures. Namely, the aim of motif scanning applications is to detect as many significant motifs as possible - these are the {\em true positives}, while keeping the number of wrong assignments - or {\em false positives} - to the minimum. In other words, accuracy is measured by how closely the response matches the set of biologically relevant sequences in the sample. 
  
Typically, motif scanning methods are based on a scoring function - usually a log-likelihood ratio - and the response is generated in two steps: first, by ranking all elements of the sample by their similarity to the query, and, second, by considering all candidates with a similarity score above a certain threshold. Both steps can be  a source of errors - an inaccurate ranking scheme, with a low threshold, will generate a huge response, causing a large type I-like error, whereas a high threshold, while testing positive on strongest examples,  might miss the candidates with a slightly weaker signal. 
Various applications deal with these problems in various ways - for example, JackHMMer (\cite{Hmmer}) uses  - effectively - several scoring functions, each with its own ranking and threshold, while IGLOSS (\cite{Pavle2019}) uses detailed parameter estimation to minimize both issues.

In this paper, we explore an alternative approach for improving accuracy, based on pairwise similarity. Namely, given a large response, containing - presumably - a large percentage of false positives, we search for subsets in which each pair of elements is sufficiently similar. We then consider the largest of these subsets as the new, modified response.  Since two true positives are more likely to be similar than a true and a false positive or two false positives, this is a sensible strategy, and it also turns out to be very robust. Furthermore, we determine what ``sufficiently similar" means - that is, we compute the appropriate similarity threshold, in terms of the expected conservation and length of the motif. The algorithm is presented in the framework of graph theory, where the new response is obtained as the maximal clique in a certain graph, that is, in turn, derived from the response graph, modified by applying the similarity threshold. As already mentioned, using the maximal clique as the new response greatly increases accuracy of the search.

\section{Methods}

\subsection{Response Graph, Derived Graph and Maximal Clique Algorithm}
\lb{rg}
The starting point for our analysis is a set of positives from a motif-scanning process. We assume that our motifs are ungapped, so this set, called response, is just a collection of k-mers. We form an undirected, weighted graph $\Gamma = (V, E )$, with vertices given by the elements of the response, and weights $w(e,f)$, $e, f \in V$, given by
\begin{jba}
w(e,f) = \sum_{i=1}^k B(e_i , f_i),
\end{jba}where $B( \cdot , \cdot )$ is the BLOSUM50-score for corresponding amino acids. 

We now form an undirected $\{ 0,1 \}$-graph $\bar{\Gamma} = (V , \bar{E}) $ using the expected weight $st$ from Section \ref{st}: for $e,f \in V$ and the unordered pair $\{ e , f \}$ we have
\begin{jba}
\{ e,f \} \in \bar{E} \Leftrightarrow w(e,f) - st > 0.
\end{jba}Put simply, two vertices in $\bar{\Gamma}$ are connected if and only if their similarity score is above the threshold. 

It is now straightforward to apply a maximal-clique algorithm to $\bar{\Gamma}$. We used the standard Bron-Kerbosh algorithm (\cite{BKClique}), implemented in Python.

\subsection{Accuracy Measures}

In this subsection, we establish notation and define relevant accuracy measures, and we closely follow  (\cite[Supplementary material, Section 5]{Pavle2019}) in the presentation. Sequences in the sample that have been annotated as belonging to test families are marked as {\em condition positive} and their number is denoted as $|CP|$, while the rest are marked as {\em condition negative} (CN). Now, each application or combination of applications under consideration produces - for a specified similarity level - a list of hits, and their respective sequences are denoted as {\em positive} (P) - with the rest of the sample being {\em negative} (N) - while $|P|$ and $|N|$ denote the corresponding set sizes.  We then  have {\em true positives} (TP) and {\em false positives} (FP) as 
\begin{jbas}
TP=P \cap CP, \ FP = P \cap CN,
\end{jbas}and likewise for {\em true negatives} (TN) and {\em false negatives} (FN)
\begin{jbas}  
TN=N \cap CN, \ FN = N \cap CP.
\end{jbas}

The usual way of assessing  diagnostic ability of an application would be to compare {\em sensitivity} or {\em true positive rate} 
\begin{jbas}
TPR = |TP|/|CP|
\end{jbas}and {\em false positive rate} 
\begin{jbas}
FPR = |FP| / |CN|. 
\end{jbas}However, in the present context, there is a serious imbalance between the sizes of the condition positive and condition negative sets: CN is several orders of magnitude larger than CP, so, for any reasonable test outcome, $|FPR|$ will be close to $0$. Consequently, we consider {\em precision} or {\em positive predictive value}
\begin{jbas}
PPV = |TP|/|P|,
\end{jbas}and use $PPV$ and $TPR$  as accuracy measures. Finally, we combine these two by considering  their harmonic mean, called the F1-score, hence
\begin{jbas}
F1=2 \cdot \frac{PPV \cdot TPR}{PPV+TPR},
\end{jbas}and plot threshold-F1 diagrams to assess accuracy.

\subsection{Similarity Threshold}
\lb{st} 
We compute the similarity threshold as the expected BLOSUM (\cite{blosum}) score of two $k$-mers, sampled from a certain distribution. Namely, given $x=( x_1 , \ldots , x_k )$ and $y = (y_1 , \ldots, y_k )$, let
\begin{jba}
s(x,y) = \sum_i B(x_i , y_i )
\end{jba}be their BLOSUM score and $\E [ s(x,y)]$ the expected value. Then, somewhat informally,
\begin{jba}
\E [ s(x,y)] = \sum \E [B(x_i , y_i ) ],
\end{jba}and, averaging over the whole length of the motif, 
\begin{jba}
\E [ s(x,y)] = k  \E [B(x_0 , y_0 ) ],
\end{jba}for some ``average" amino acids $x_0$ and $y_0$. Now, let 
\begin{jba}
e_i = (0, \ldots , 1, 0, \ldots, 0),
\end{jba}with $1$ on $i$-th position, let $\alpha \in (0,1)$, and let $bg=bg(i)$ be the average distribution of amino acids. Let
\begin{jbas}
\lb{eq1}
f_i = \alpha \cdot e_i + (1-\alpha) bg
\end{jbas}be an $e_i - bg$-mixture of distributions. Here, $\alpha$ is a ``conservation parameter", representing the percentage of the dominant amino acid $i$ in an alignment column sampled from $f_i$. Then
\begin{jba}
\sum_{j,k=1}^{20} B(j,k) f_i (j) f_i (k)
\end{jba}is the expected BLOSUM score for two amino acids sampled from $f_i$. Setting $\alpha = 0.68$ and averaging over the distribution $bg$, we get
\begin{jbas}
\lb{eq2}
\sum_{i=1}^{20}bg(i)\sum_{j,k=1}^{20}B(j,k)f_i(j)f_i(k)=2.522,
\end{jbas}and, for a motif length $k$, we take the similarity threshold $st$ to be
\begin{jbas}
st = k \cdot 2.5
\end{jbas}We further comment on this in Section \ref{simthresh}.

\section{Results and Discussion}

\subsection{Tests and Results}
\lb{tar}

In order to test the method, we applied it to responses from three iterative motif scanners - PSI-BLAST (PB) (\cite{Altschul1997}), JackHMMER (JH) (\cite{Hmmer}) and IGLOSS (IG) (\cite{Pavle2019}) - and compared the maximal clique with the original response. As in (\cite{Pavle2019}), scanners were applied to five plant proteomes - Arabidopsis thaliana (AT, v. TAIR9), Oryza sativa (OS, v. MSU v7), Solanum tuberosum (ST, v. ITAG1), Solanum lycopersicum (SL, v. ITAG2.3) and Beta vulgaris (BV, v. KWS2320) - where we searched for members of an extensively studied, motif characterized protein family - GDSL lipases. 

GDSL lipases belong to lipid hydrolyzing enzymes that exhibit a GDSL motif. Proteins in this family display fairly low overall sequence similarity, but are reasonably well described by the presence of
conserved residues in four conserved blocks (I, II, III, and V) (\cite{Vujaklija2016}). Block I contains the
main characteristic motif (PROSITE:PS01098) (\cite{Sigrist2013}) from which the main search query of 10 amino acids was constructed. As in \cite{Pavle2019}, the condition positive set was determined by processing the information from GoMapMan resource \cite{Ramsak2014}.

Altogether, we performed approximately $900$ tests. We used three search queries:
\begin{itemize}
\item FVFGDSLSDA - consensus query, defined above
\item FVFnDSLSDA - a single mutation, at a highly conserved site
\item vfFGDSLSDn - three substitutions
\end{itemize}
This was done with all three scanners, for approximately $20$ threshold levels each, and all five proteomes.

The average gain in the F1-score was around $0.20$ in the most interesting threshold region (around 1/3 of the x-axis). However, gains can sometimes be spectacular - Table \ref{table:table4} shows results of scans where we more than doubled F1-score, albeit at a fairly low threshold level. Another feature of our tests that we should comment on are peaks-and-troughs present in the Figure \ref{fig:PB-3}. Namely, the corresponding responses are diffuse and inhomogeneous, as a result of a ``wrong'' query, so the clique algorithm hardly improves PPV at certain threshold levels - hence a ``jagged'' PPV-curve. We further analyzed this situation, and detected features that indicate that the clique recognizes this ``wrongness''. We discuss this in Section \ref{pat}.

Figures in the last section were obtained by merging the results from all five proteomes. Here, responses are matched by their size (i.e. the number of positives), and the average threshold - scale (for IG) or negative logarithm of the e-value (for PB and JH) - was assigned to the x-axis, with averaged PPV, TPR or F1-score on the y-axis. For the tables, responses were again matched by their sizes, with the last column reporting cumulative results for positives and true positives, and the average threshold.

\subsection{Discussion}
\subsubsection{Overview}
Put somewhat abstractly, the aim of our method is to detect - using pairwise comparison - an optimal subset of true positives in a (fairly) large set of hits (positives). This is achieved with the help of a pairwise similarity threshold and a maximal clique algorithm. We have developed this concept in the context of ungapped motif scanning - hence the present method - and tested it in conjunction with three different iterative motif scanners. As stated before, these principles can be applied more generally, provided suitable assumptions are fulfilled. 
\subsubsection{Robustness}
\begin{figure}[!h]
	\centering
	\begin{subfigure}{.45\textwidth}
		\centering
		\includegraphics[width=6cm]{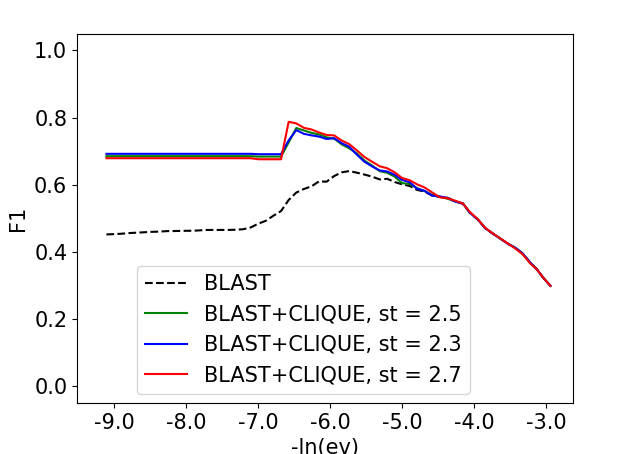}
		\caption{robustness BLAST}
		\label{fig:blastf1}
	\end{subfigure}
	\begin{subfigure}{.45\textwidth}
		\centering
		\includegraphics[width=6cm]{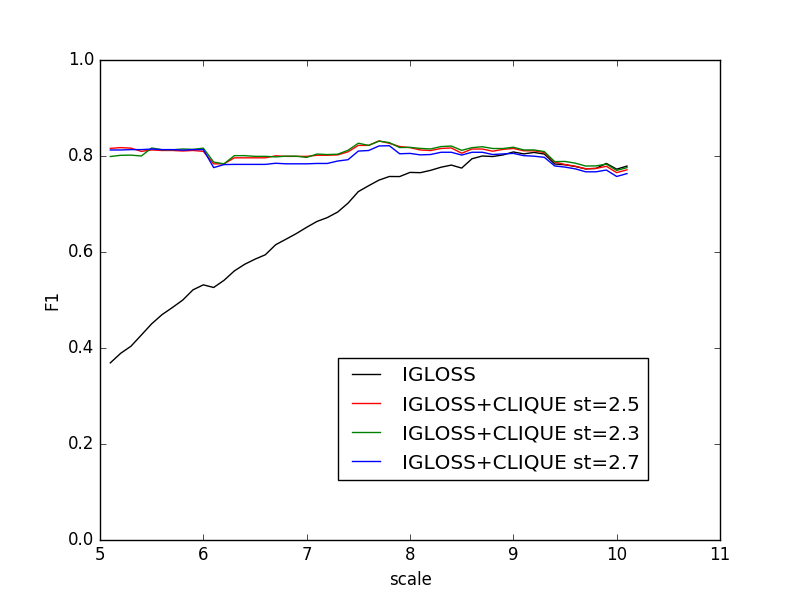}
		\caption{robustness IGLOSS}
		\label{fig:os}
	\end{subfigure}
	\caption{F1-threshold level}
\end{figure}

Before we discuss the way we computed the similarity threshold - in the next section - let us first note that the threshold is rather robust. Namely, as seen from the Figure \ref{fig:blastf1}, $10\%$ change in the level of the threshold produces only minor changes in terms of accuracy (measured by the F1-score). Hence, gradual changes in the threshold level will affect accuracy gradually. On the other hand, this is in contrast to the iterative approach - as mentioned in the Introduction - where small changes of the threshold (e-value or scale) might produce very different results. This is due to the different nature of two approaches -  there are many more comparisons carried out in the pairwise case, so the effect of the change is smoothed out. Furthermore, to phrase this in the graph theory framework: the pairwise threshold acts on the edges of the response graph, while we are looking for a suitable subset of the vertices; so, while a small change in the threshold level might affect many edges, that will - eventually - add or remove only a couple of vertices from the maximal clique.    
\subsubsection{Similarity Threshold Comments}
\lb{simthresh}
The aim of the similarity threshold is to distinguish between true and false positives by means of the pairwise score. Namely, protein motifs are chracterized by a specific substitution pattern, so it is to be expected that the pairwise score between two true positives will be higher than the one between true and false positives, let alone two false positives (note that this naturally leads to the maximal clique approach). However, knowledge of the substitution pattern amounts to a detailed description of the motif under consideration, in which case the type of analysis that we study here - single-query, iterative approximation - becomes almost redundant. 

Hence, we derived the threshold {\em a priori}, as an abstract, average score, and dependent on a single parameter $\alpha$, so-called {\em conservation coefficient}. More precisely, the threshold was obtained as the expected BLOSUM score of two $k$-mers, sampled from an $\alpha$-convex combination of distributions, as well as assuming average (i.e. background) distribution of amino acids across the length $k$. The parameter $\alpha$ should be thought of as the relative frequency of the dominant amino acid in the hypothetical motif profile, averaged across the length $k$. Incidently, in five proteomes that we considered, profile conservation varied from $68\%$ to $72\%$. 

There are other strategies to obtain the threshold, as long as appropriate conservation is maintained. For example, one could use a suitable power of the PAM matrix (\cite{Dayhoff1978}) instead of distributions $f_i$ in the Equation \ref{eq1}. So, taking PAM120 - which yields the average diagonal value around $2/3$ - and repeating the procedure from Section \ref{st}, gives the value $st=2.58$ - approximately the same. Likewise, using uniform distribution - instead of background - in the Equation \ref{eq2}, one obtains $st=2.4$. Furthermore, other similarity measures, rather than the BLOSUM50 matrix, could be used - other BLOSUM matrices, or even other, non-BLOSUM substitution matrices. However, that would involve setting up a new response graph and a new threshold, in parallel. Considering the comments above, one should expect marginal changes, or no changes at all. 

Finally, we should mention that, although the threshold was obtained as an expected score, it is used as the ``minimal allowed" score. This is because  assumptions for the conservation and, especially, amino acid composition are  rather weak. It is possible to tighten these assumptions, and then set the ``minimal score" to, say, $-2\sigma$ (i.e. two standard deviations) from the mean. However, tightening the assumptions would again amount to a description of the motif under consideration - hence, not an {\em a priori} approach - so we decided not to explore this further.       

\subsubsection{Peaks-and-Troughs}
\begin{figure}[!h]
	\includegraphics[width=5cm]{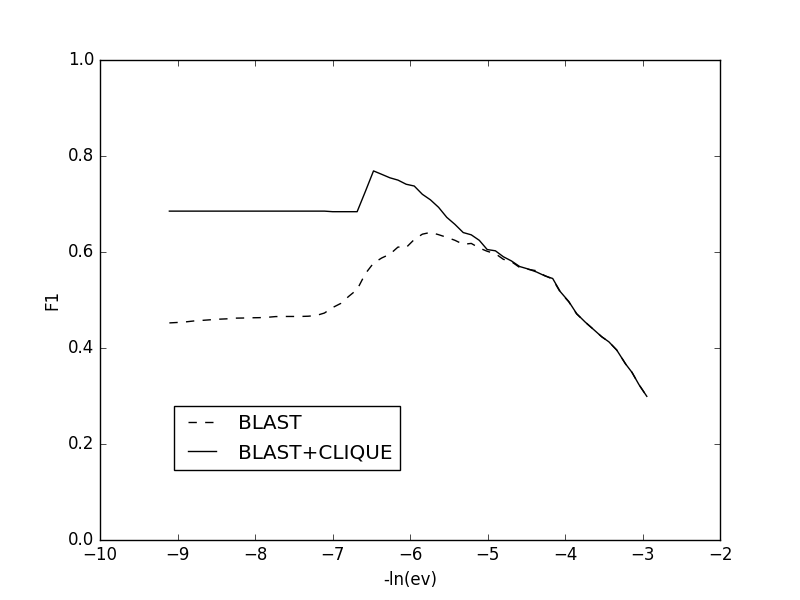}
	\includegraphics[width=5cm]{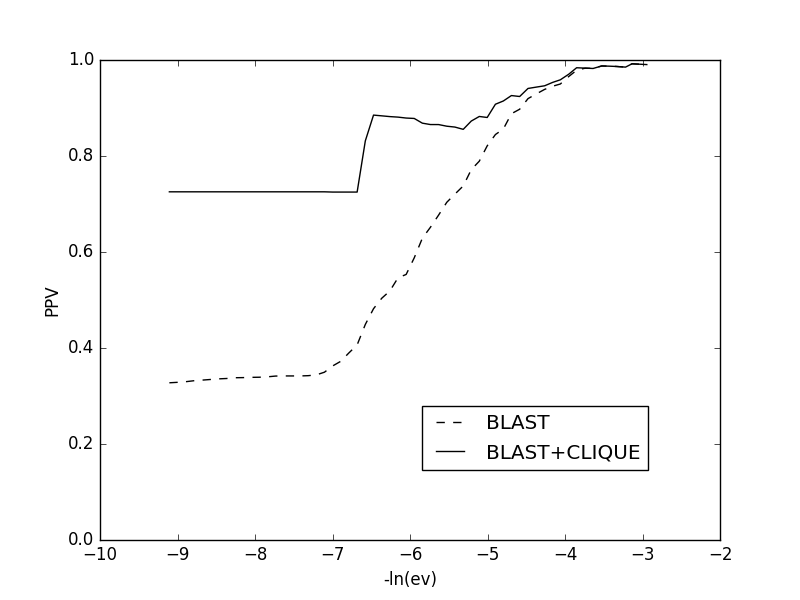}
	\includegraphics[width=5cm]{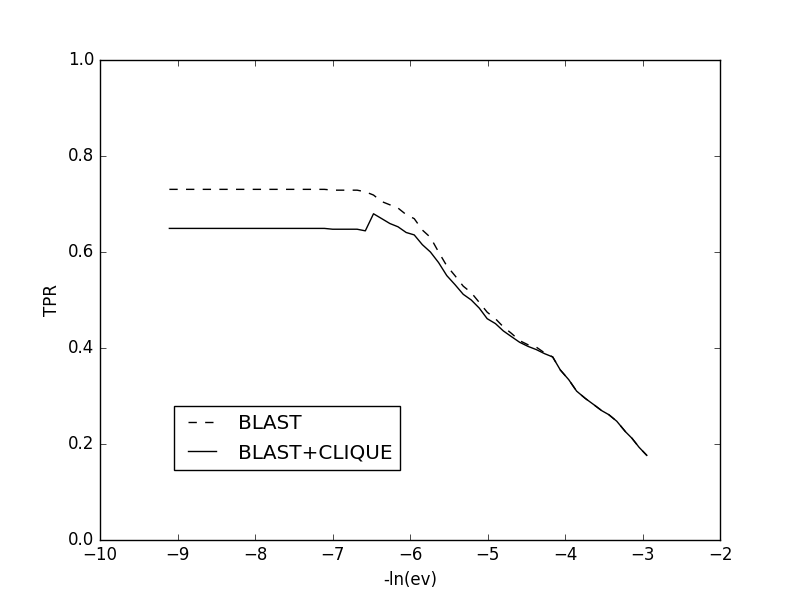}

	\caption{BLAST FVFGDSLSDA}
	\label{fig:PB-1}
\end{figure}

\begin{figure}[!h]
	\includegraphics[width=5cm]{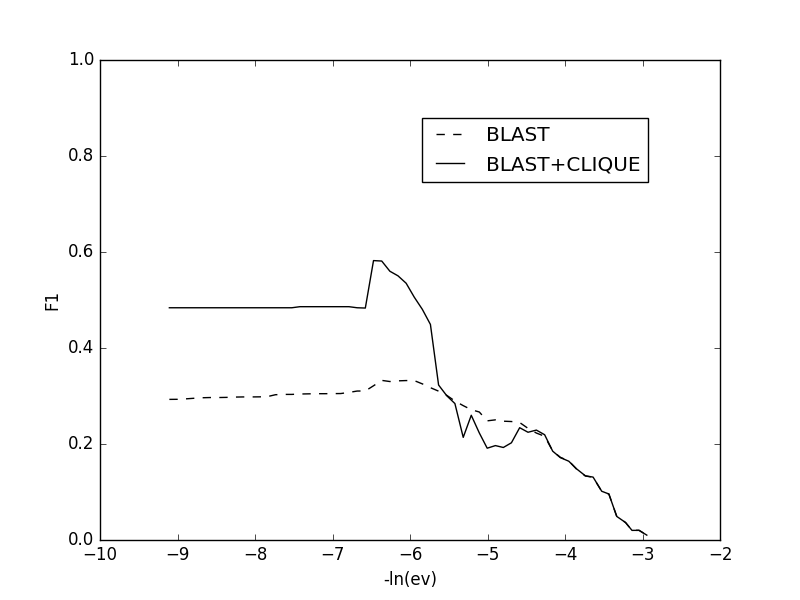}
	\includegraphics[width=5cm]{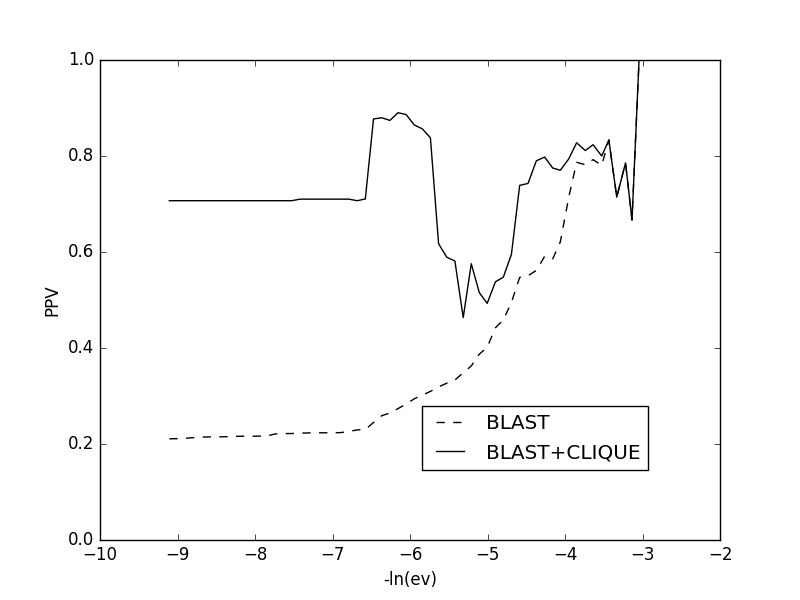}
	\includegraphics[width=5cm]{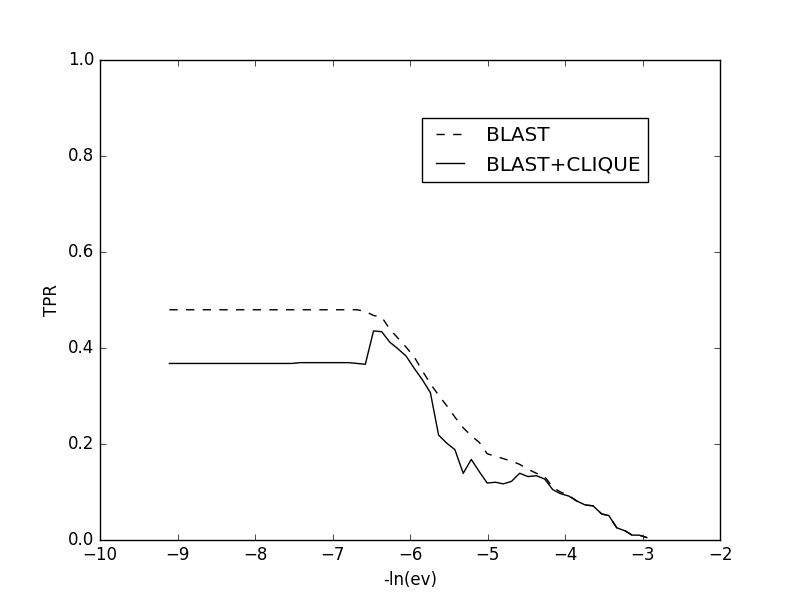}

	\caption{BLAST VFFGDSLSDN}
	\label{fig:PB-3}
\end{figure}

\lb{pat}
As mentioned in Section \ref{pat}, peaks-and-troughs in the Figure \ref{fig:PB-3} were caused by the incorrect query, resulting in an inhomogenous, diffuse response. This, in turn, produced the maximal clique containing a large number of false positives, yielding a fairly low PPV for some threshold levels. Obviously, this is in contrast with Figure \ref{fig:PB-1}, where the ``right'' - i.e. consensus - query produces more homogenous response, and the clique yields consistent improvement. We analyzed this a bit futher, in order to detect features that might differentiate between these two situations. 

First of all, note that most of the scan descriptions used above - ``low PPV, varying F1-score'', and figures such as Figure \ref{fig:PB-3} and Figure \ref{fig:PB-1} - are available only {\em a posteriori}. In other words, this information is not available in an exploratory setting, where one wants to asses the validity of the response without knowing the desired outcome. So, we should be looking for some other - {\em ``a priori''} - features. 

Homogeneity of the response is a possible candidate, and it will distinguish between these two sets of scans, but, again, homogeneity is a feature that is best measured {\em a posteriori}, when some information regarding the variability of the condition positive set is available. So, we looked at {\em stability}, that is, relationship between responses and their respective cliques, for neighbouring threshold levels. 

A sequence of threshold levels, from low to high, should - in principle - produce a descending family of responses. This is, clearly, the case in a non-iterative setting, where a simple scan results in a fixed ranking scheme - an ordering of the sample with respect to the similarity to the query. In an iterative situation, the similarity function is being optimized, which might produce a different ranking scheme from one iteration to another, and result in a different ranking scheme from one threshold to another. Consequently, a smaller response - produced with a higher threshold - might not be the subset of a larger one, obtained with a lower threshold. Furthermore, a significant deviation from this stability principle indicates, in general, problems with either the sample or the query.

Let us analyze from this point of view the series of scans from Figure \ref{fig:PB-3} (we will focus on Arabidopsis thaliana here; some of the scan results - for neighbouring e-values - are presented in Tables \ref{table:table3} and \ref{table:table1}; for a complete set of tables, consult the server web-page): as already mentioned, we have scanned with a non-consensus query; quite surprisingly, responses have shown to be rather stable, with sizes $98$ and $81$, and the size of the intersection $80$; however, the corresponding cliques - with sizes $26$ and $24$ - have a single element in the intersection, which should be considered as a significant deviation from the stability principle. Hence, our approach provides another method to assess validity of a scan, with the clique as a new stability criterion.

\subsubsection{Pairwise Similarity vs EM-algorithm}

All the iterative motif scanners that we combined our clique-method with use some form of the expectation maximization algorithm (EM-algorthm) to find the ``optimum" - the optimal set of positives, for a given significance level. On the other hand, the maximal clique algorithm also provides the optimal solution for the set of positives - the maximal clique itself. A natural question arises: are these two optima the same? More precisely, given the right parameters - similarity and significance threshold - will these two approaches return the same, or very similar, response?

The answer appears to be yes - optimal solutions will be more-or-less the same for the right choice of parameters, with a couple of outliers added or subtracted. This can already be inferred from figures in the next Section, where we see stability in TPR across all threshold ranges, and F1-convergence, as the threshold becomes higher. In the opposite direction, iterative scanners will, invariably, accept the maximal clique as the optimal solution, again with minor changes. 

All this means that these two approaches are complementary and interchangeable, at least in the present context. How general can such framework be is, at the moment, unclear. The underlying algorithms - the EM-algorithm and maximal-clique approach - are very different, so this agreement should not be considered a rule. However, in a very tractable situation - an $n$-dimensional Euclidean space $\E^n$- this can be made more precise, as follows: we have to show that sets of positives from the two approaches - the EM-algorithm and the maximal clique - are identical; it is well known that in $\E^n$ the k-means clustering algorithm is a form of EM-algorithm (see \cite{esl}), with clusters (i.e. sets of positives) given by $n$-balls, where
\begin{jbas}
B(x_0 ; r)= \{ x \in \E^n ; |x - x_0 | \leq r \}
\end{jbas}is an $n$-ball around the point $x_0$, with the radius $r$; on the other hand, set $d=2r$, and note that $B=B(x_0 ; r )$ can be realized as the largest subset of $\E^n$ such that
\begin{jbas}
B=\{ x \in \E^n ; |x-y| \leq d, \forall y \in B \},
\end{jbas}which yields a clique-like object.

Consequently, we see that the considerable improvement in accuracy that we have recorded is not a question of a superior, but complementary approach. Namely, our queries - deliberately - consist of a single string, sometimes even not a consensus query, and iterative process reaches a local optimum - a set of positives with a fairly low F1-score and a blurred signal. And this is a situation where maximal clique approach offers greatest gains, by filtering the response and providing a better foundation for the next step in analysis.

\subsubsection{Availability}

\url{http://compbioserv.math.hr/igloss/index.html?clique}
\section{Tables and Figures}

\subsection{BLAST}
\begin{figure}[!h]
	\includegraphics[width=5cm]{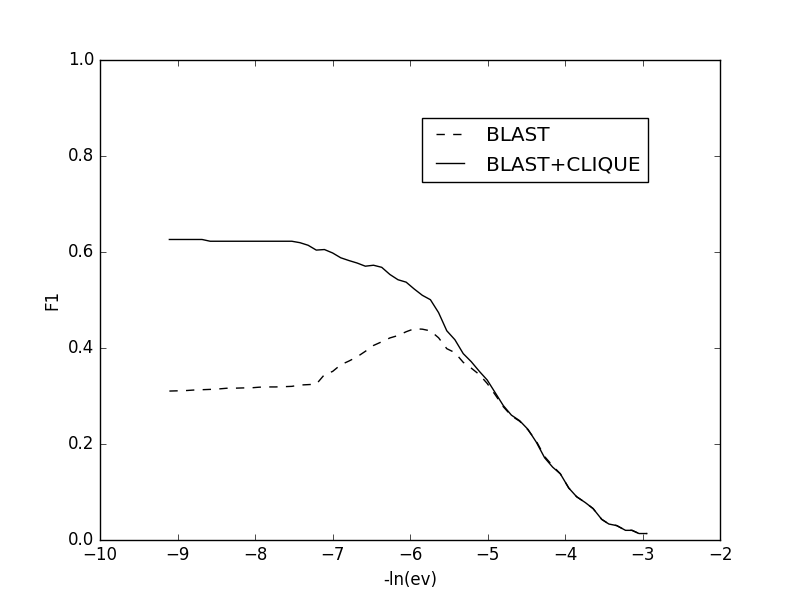}
	\includegraphics[width=5cm]{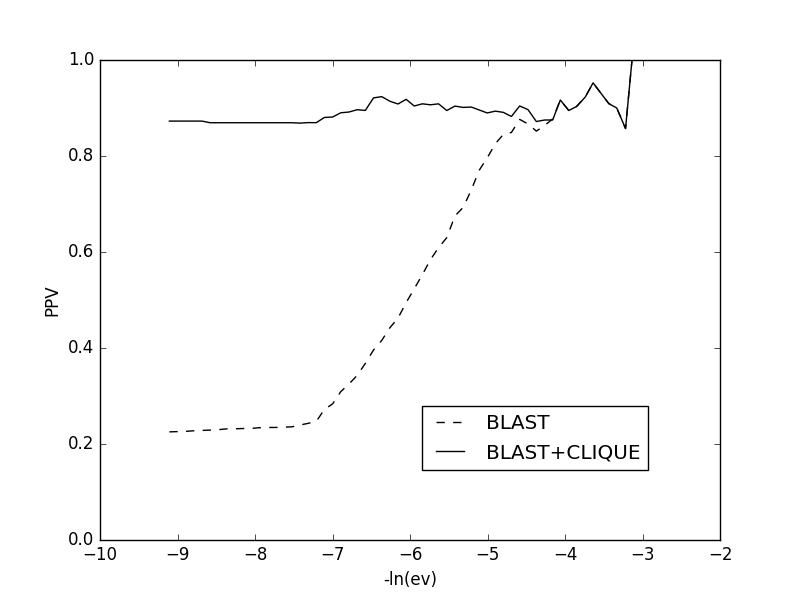}
	\includegraphics[width=5cm]{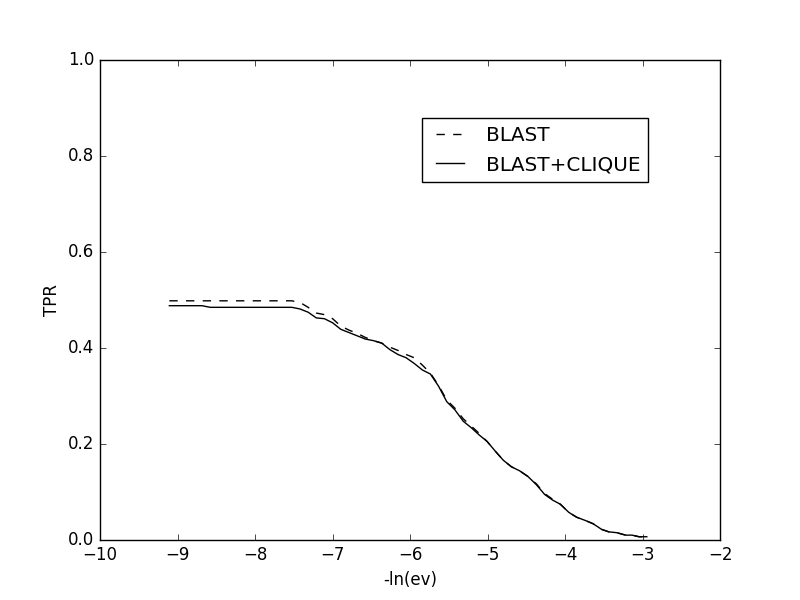}

	\caption{BLAST FVFNDSLSDA}
	\label{fig:PB-2}
\end{figure}

\FloatBarrier
\subsection{JackHMMER}

\begin{figure}[!h]
	\includegraphics[width=5cm]{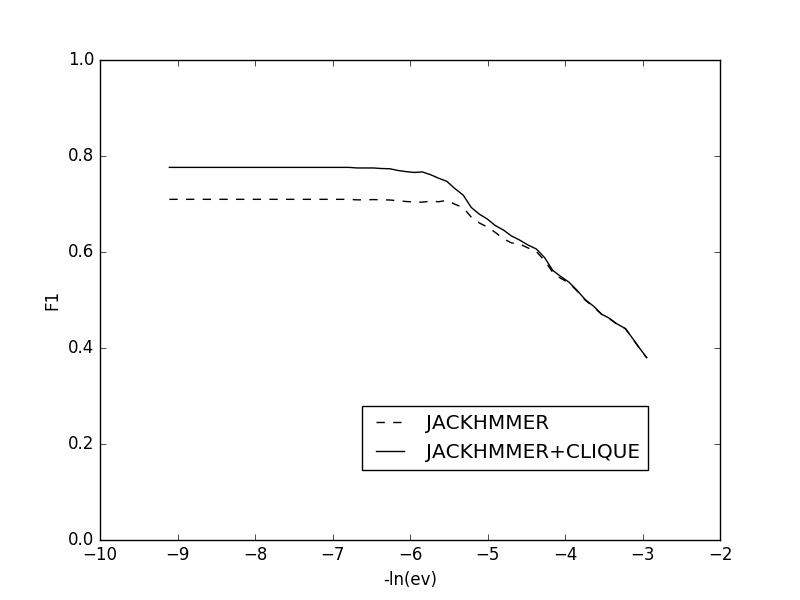}
	\includegraphics[width=5cm]{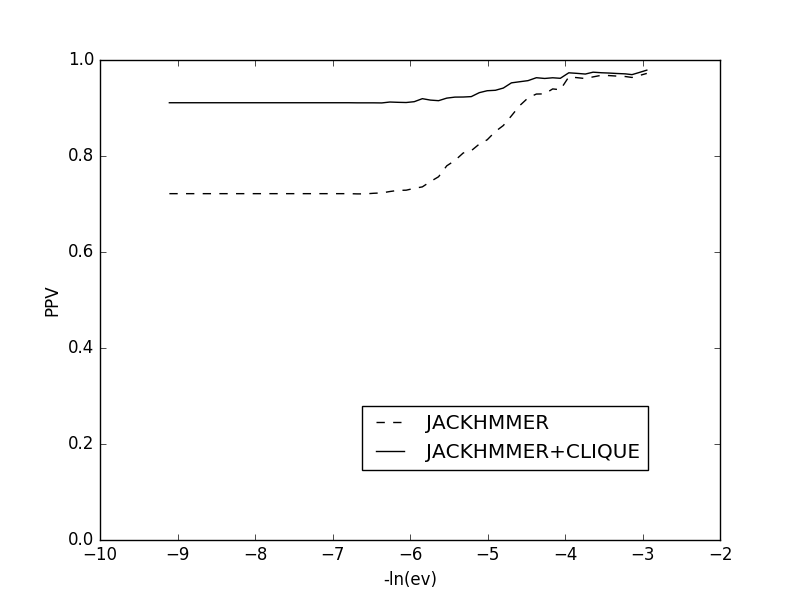}
	\includegraphics[width=5cm]{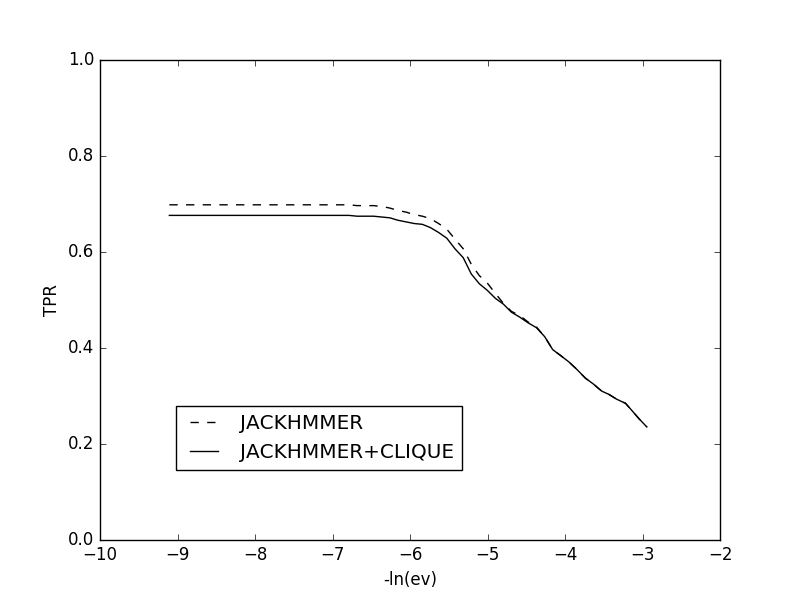}

	\caption{HMMER FVFGDSLSDA}
	\label{fig:JH-1}
\end{figure}
\begin{figure}[!h]
	\includegraphics[width=5cm]{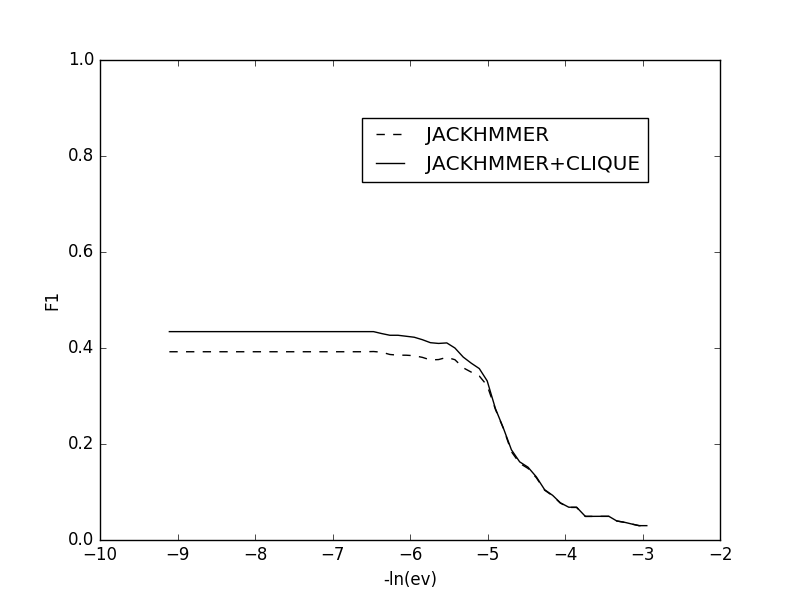}
	\includegraphics[width=5cm]{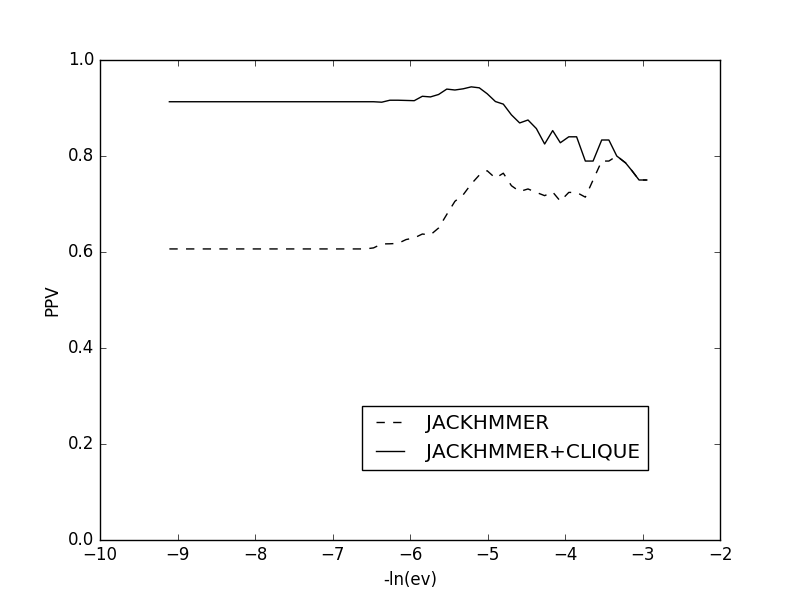}
	\includegraphics[width=5cm]{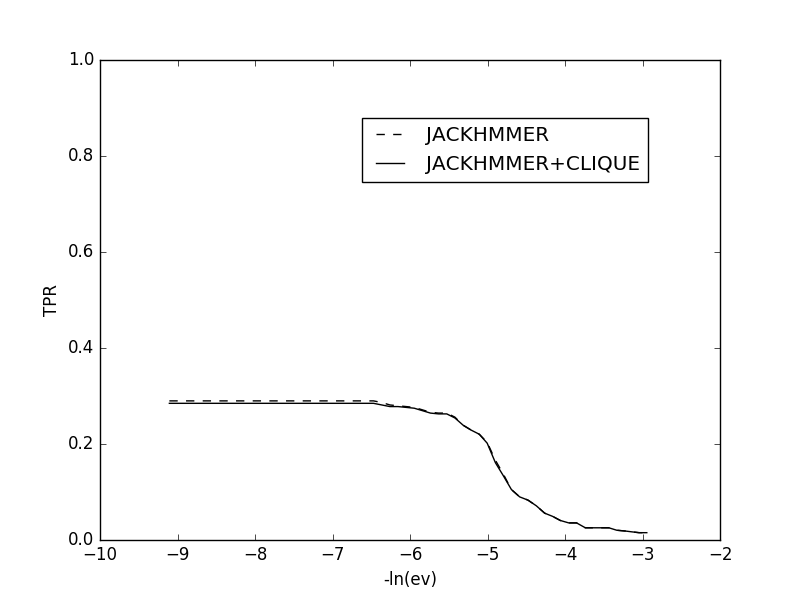}

	\caption{HMMER FVFNDSLSDA}
	\label{fig:JH-2}
\end{figure}
\begin{figure}[!h]
	\includegraphics[width=5cm]{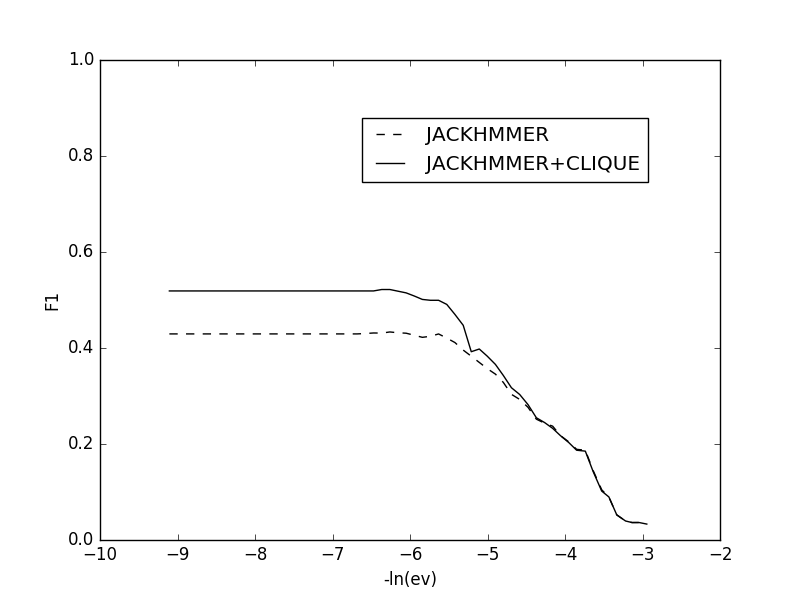}
	\includegraphics[width=5cm]{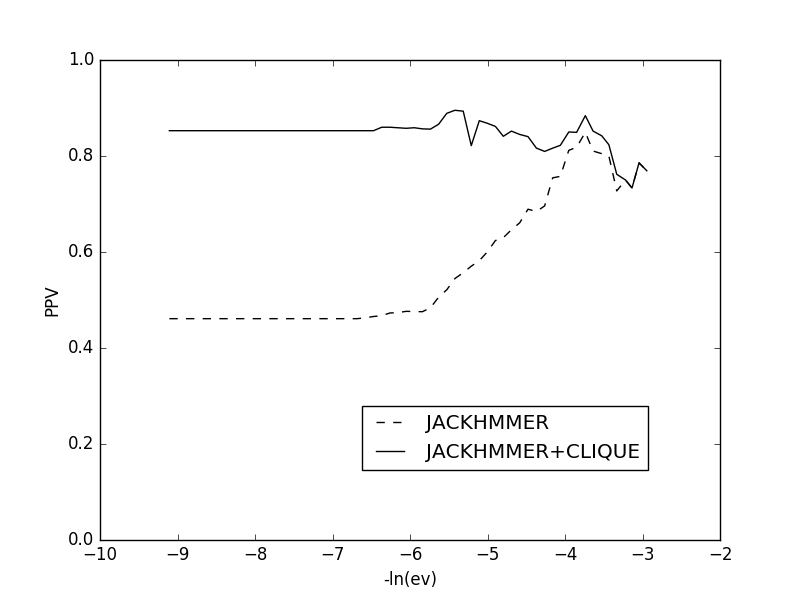}
	\includegraphics[width=5cm]{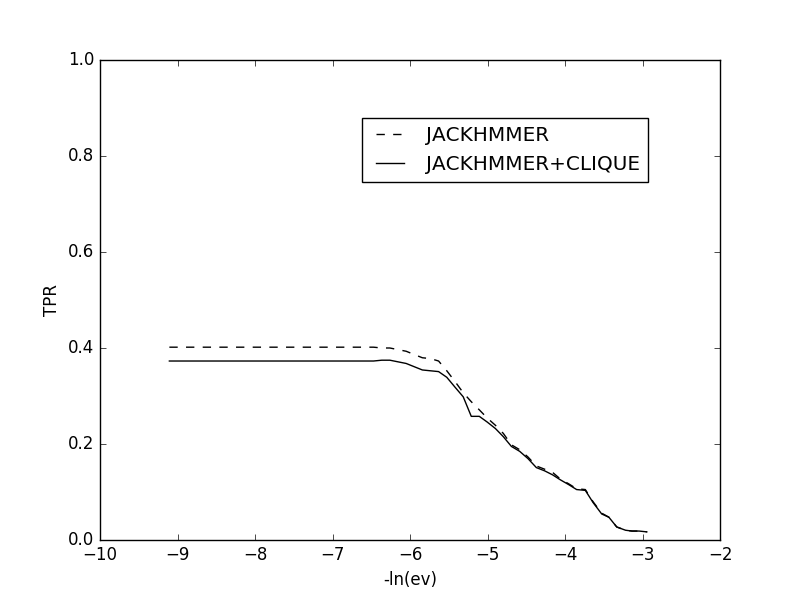}

	\caption{HMMER VFFGDSLSDN}
	\label{fig:JH-3}
\end{figure}

\newpage
\subsection{IGLOSS}

\begin{figure}[!h]
	\includegraphics[width=5cm]{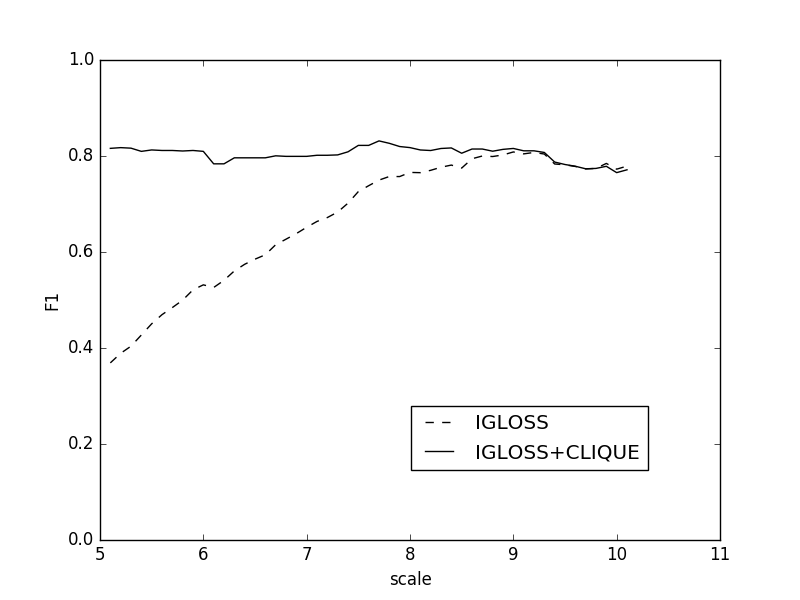}
	\includegraphics[width=5cm]{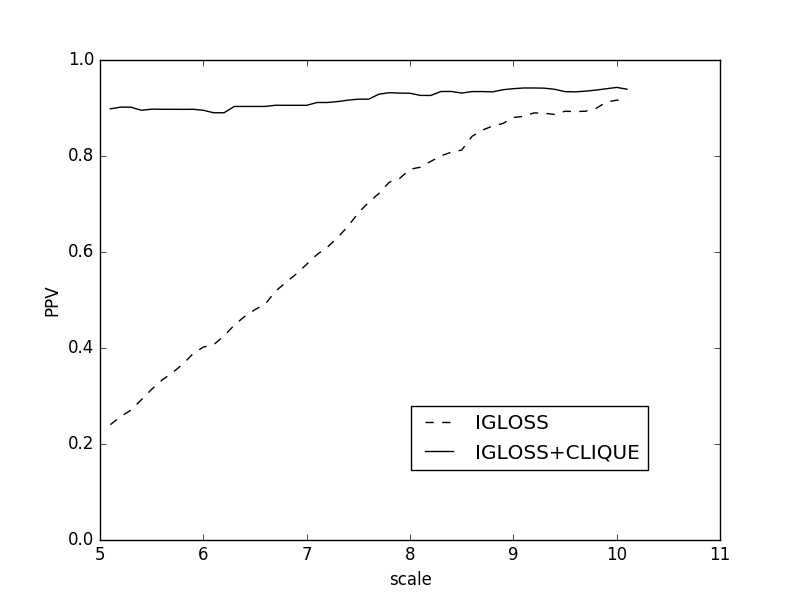}
	\includegraphics[width=5cm]{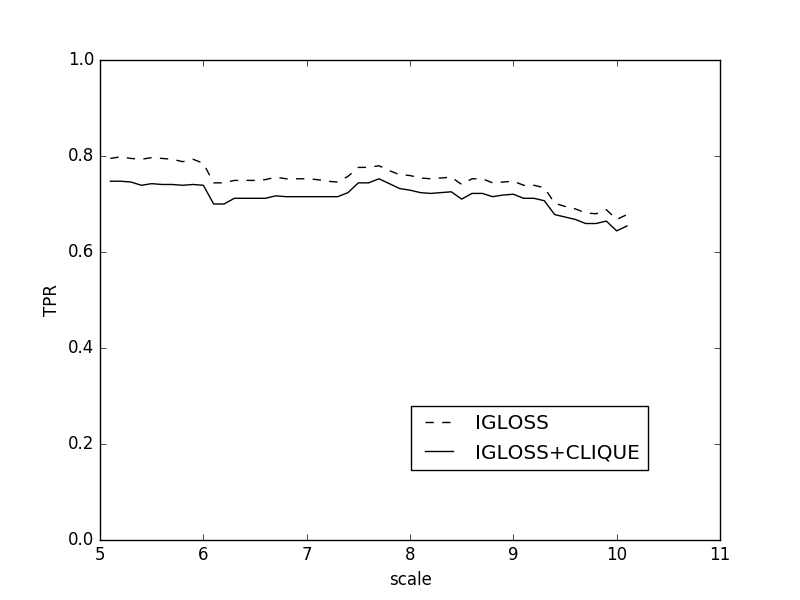}

	\caption{IGLOSS FVFGDSLSDA}
	\label{fig:IG-1}
\end{figure}
\begin{figure}[!h]
	\includegraphics[width=5cm]{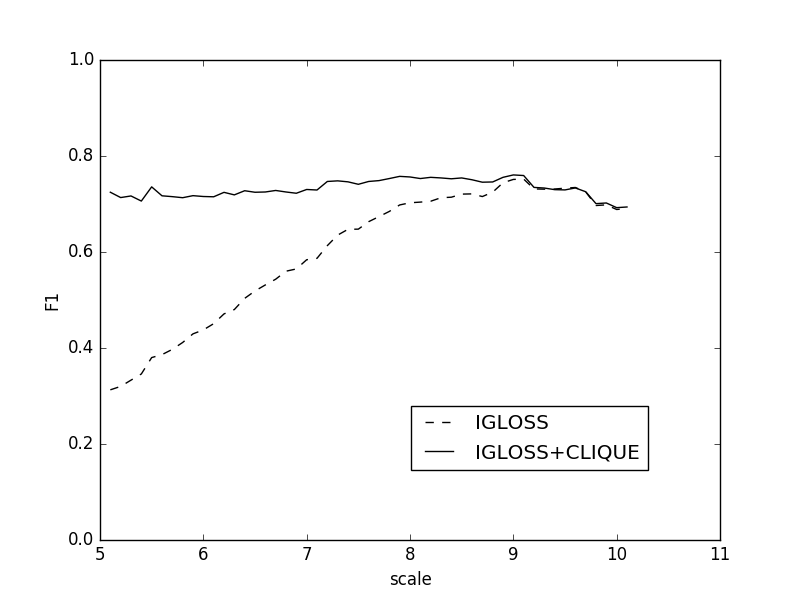}
	\includegraphics[width=5cm]{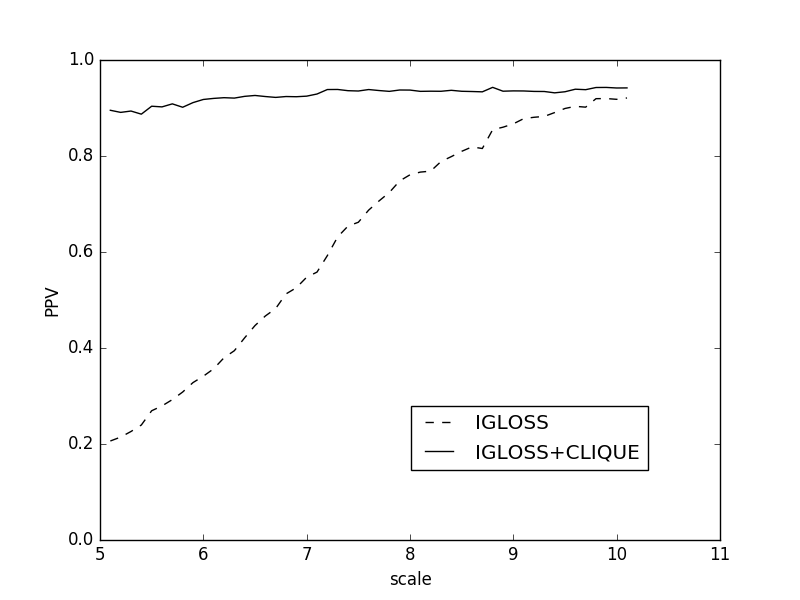}
	\includegraphics[width=5cm]{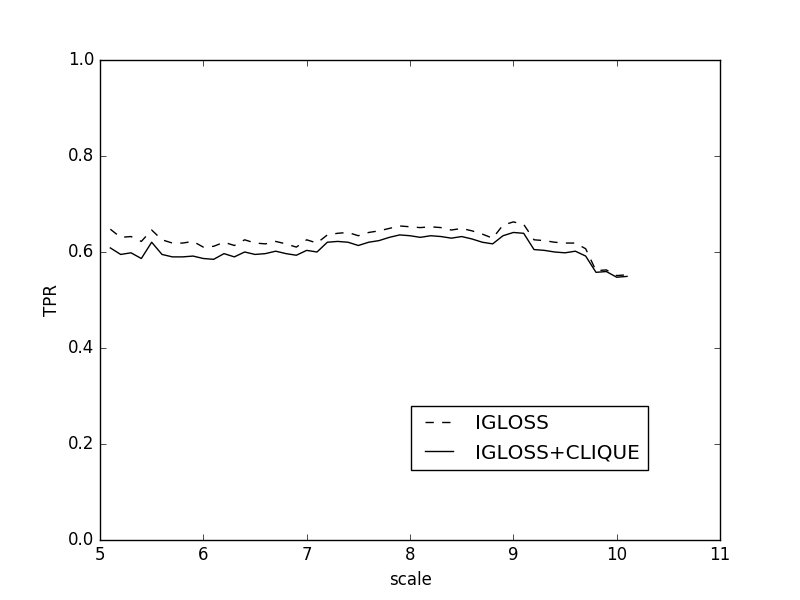}

	\caption{IGLOSS FVFNDSLSDA}
	\label{fig:IG-2}
\end{figure}
\begin{figure}[!h]
	\includegraphics[width=5cm]{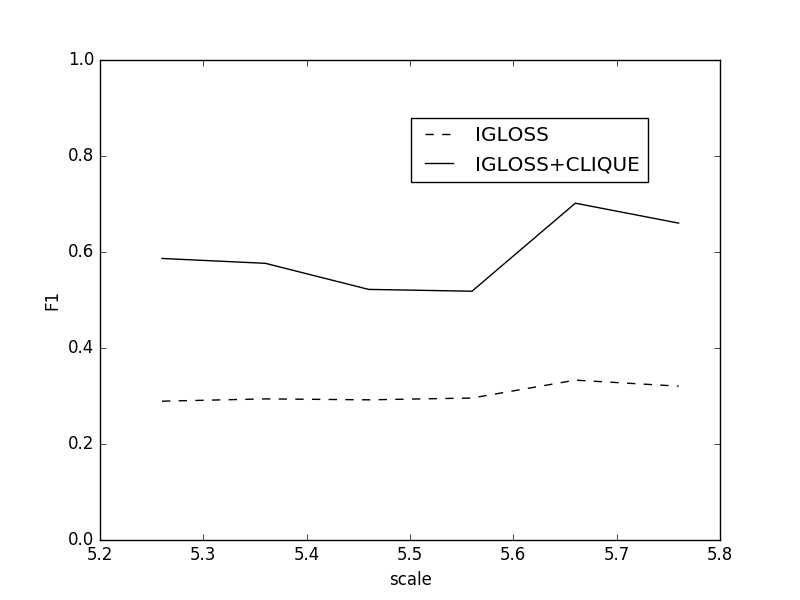}
	\includegraphics[width=5cm]{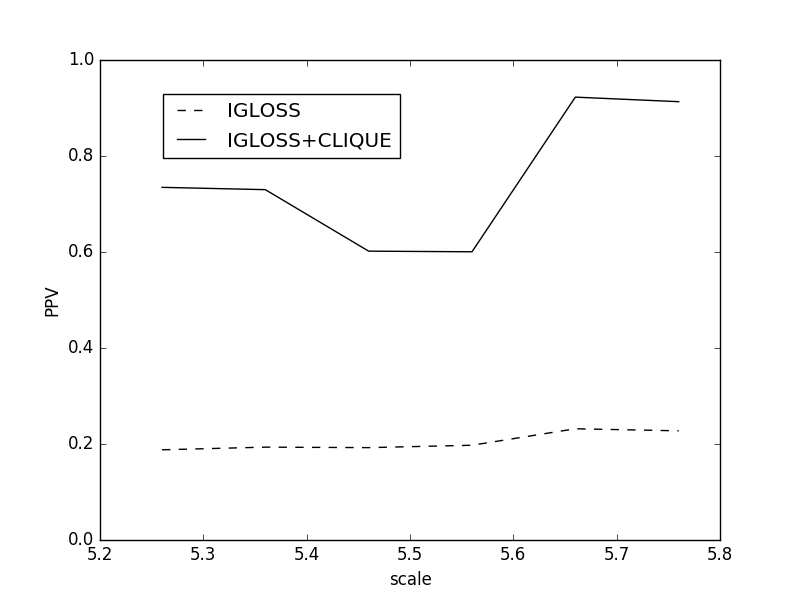}
	\includegraphics[width=5cm]{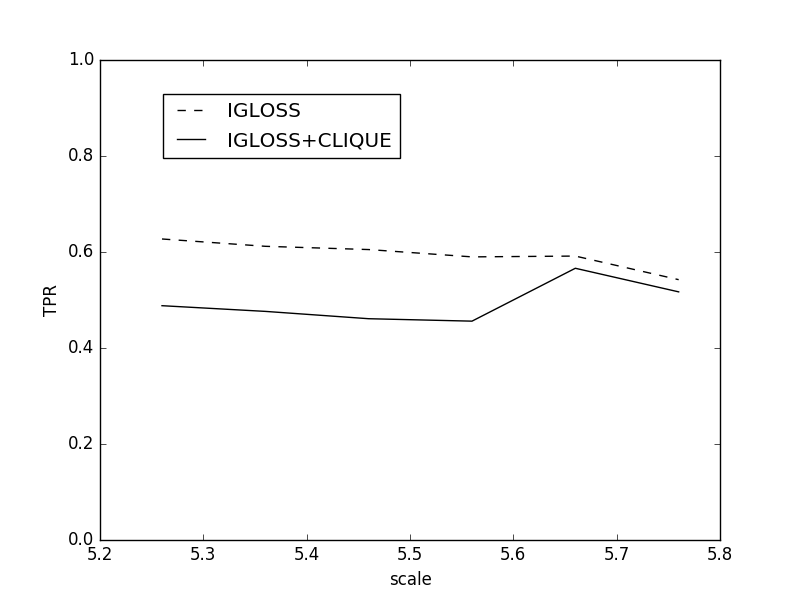}

	\caption{IGLOSS VFFGDSLSDN}
	\label{fig:IG-3}
\end{figure}

\newpage
\subsection{Examples}

\subsection{Tables}

\begin{table}[!h]
	\begin{center}
		
		\begin{tabular}{|c|c|c|c|c|c||c|}
			\hline
				         &     AT    &     OS    &     ST    &     SL    &     BV    &   ALL    \\
			\hline
			CP      &       118 &       155 &       123 &       108 &        86 &       590\\
			\hline
			& \multicolumn{5}{|c||}{IGLOSS} &\\
			\hline
			scale   &       5.0 &       6.0 &       5.0 &       4.9 &       4.6 &       5.1\\
			TP/P    &  105/421 &  111/321 &   95/389 &   94/408 &   64/413 &  469/1952\\
			PPV     & 0.2494 & 0.3458 & 0.2442 & 0.2304 & 0.1550 & 0.2403\\
			TPR     & 0.8898 & 0.7161 & 0.7724 & 0.8704 & 0.7442 & 0.7949\\
			F1      & 0.3896 & 0.4664 & 0.3711 & 0.3643 & 0.2565 & 0.3690\\
			\hline
			& \multicolumn{5}{|c||}{IGLOSS + CLIQUE} & \\
			\hline
			TP/P    &  101/115 &  106/120 &   87/97 &   87/92 &   60/67 &  441/491 \\
			PPV     & 0.8783 & 0.8833 & 0.8969 & 0.9457 & 0.8955 & 0.8982 \\
			TPR     & 0.8559 & 0.6839 & 0.7073 & 0.8056 & 0.6977 & 0.7475 \\
			F1      & 0.8670 & 0.7709 & 0.7909 & 0.8700 & 0.7843 & 0.8159 \\
			\hline
		\end{tabular}
		\caption{GDSL lipases IGLOSS vs IGLOSS+CLIQUE (FVFGDSLSDA)}
		\label{table:table4}		
	\end{center}
	
\end{table}

\begin{table}[!h]
	\begin{center}
		\begin{tabular}{|c|c|c|c|c|c||c|}
			\hline
					 &     AT    &     OS    &     ST    &     SL    &     BV    &   ALL \\
			\hline
			CP      &       118 &       155 &       123 &       108 &        86 &       590 \\
			\hline
			& \multicolumn{5}{|c||}{IGLOSS} &\\
			\hline
			scale   &       6.6 &       7.6 &       6.6 &       6.5 &       6.2 &       6.7 \\
			TP/P    &   80/180 &  118/176 &   92/174 &   93/174 &   63/156 &  446/860 \\
			PPV     & 0.4444 & 0.6705 & 0.5287 & 0.5345 & 0.4038 & 0.5186 \\
			TPR     & 0.6780 & 0.7613 & 0.7480 & 0.8611 & 0.7326 & 0.7559 \\
			F1      & 0.5369 & 0.7130 & 0.6195 & 0.6596 & 0.5207 & 0.6152 \\
			\hline
			& \multicolumn{5}{|c||}{IGLOSS + CLIQUE} & \\
			\hline
			TP/P    &   75/92 &  116/124 &   87/95 &   86/90 &   59/66 &  423/467 \\
			PPV     & 0.8152 & 0.9355 & 0.9158 & 0.9556 & 0.8939 & 0.9058 \\
			TPR     & 0.6356 & 0.7484 & 0.7073 & 0.7963 & 0.6860 & 0.7169 \\
			F1      & 0.7143 & 0.8315 & 0.7982 & 0.8687 & 0.7763 & 0.8004 \\
			\hline
			
		\end{tabular}
		\caption{GDSL lipases IGLOSS vs IGLOSS+CLIQUE (FVFGDSLSDA)}
	\end{center}
	\label{table:table2}
\end{table}

\begin{table}[!h]
	\centering
	\begin{center}
		
		\begin{tabular}{|c|c|c|c|c|c||c|}
			\hline
				 &     AT    &     OS    &     ST    &     SL    &     BV    &   ALL \\
			\hline
			CP      &       118 &       155 &       123 &       108 &        86 &       590\\
			\hline
			 & \multicolumn{5}{|c||}{BLAST} &\\
			\hline
 ev      &       226 &       226 &       226 &       226 &       226 &       226\\
 TP/P    &   26/98 &   32/51 &   26/ 109 &   32/108 &   35/87 &  151/453\\
 PPV     & 0.2653 & 0.6275 & 0.2385 & 0.2963 & 0.4023 & 0.3333\\
 TPR     & 0.2203 & 0.2065 & 0.2114 & 0.2963 & 0.4070 & 0.2559\\
 F1      & 0.2407 & 0.3107 & 0.2241 & 0.2963 & 0.4046 & 0.2895\\
			\hline
			& \multicolumn{5}{|c||}{BLAST + CLIQUE} &\\
			\hline
			 TP/P    &   24/35 &   31/38 &   10/40 &   12/41 &   34/37 &  111/191\\
			 PPV     & 0.6857 & 0.8158 & 0.2500 & 0.2927 & 0.9189 & 0.5812\\
			 TPR     & 0.2034 & 0.2000 & 0.0813 & 0.1111 & 0.3953 & 0.1881\\
			 F1      & 0.3137 & 0.3212 & 0.1227 & 0.1611 & 0.5528 & 0.2843\\
			\hline
			
		\end{tabular}
		\caption{BLAST vs BLAST+CLIQUE (VFFGDSLSDN)}
	\label{table:table1}		
	\end{center}

\end{table}

\begin{table}[!h]
	\begin{center}
		\begin{tabular}{|c|c|c|c|c|c||c|}
			\hline
         &     AT    &     OS    &     ST    &     SL    &     BV    &   ALL \\
\hline
 CP      &       118 &       155 &       123 &       108 &        86 &       590\\
\hline
		& \multicolumn{5}{|c||}{BLAST} & \\
\hline
 ev      &       203 &       203 &       203 &       203 &       203 &       203\\
 TP/P    &   24/81 &   29/44 &   24/98 &   29/93 &   32/80 &  138/396\\
 PPV     & 0.2963 & 0.6591 & 0.2449 & 0.3118 & 0.4000 & 0.3485\\
 TPR     & 0.2034 & 0.1871 & 0.1951 & 0.2685 & 0.3721 & 0.2339\\
 F1      & 0.2412 & 0.2915 & 0.2172 & 0.2886 & 0.3855 & 0.2799\\
			\hline
	& \multicolumn{5}{|c||}{BLAST + CLIQUE} &\\
\hline
 TP/P    &    1/30 &   29/35 &   10/40 &   12/39 &   30/33 &   82/177\\
 PPV     & 0.03333 & 0.8286 & 0.2500 & 0.3077 & 0.9091 & 0.4633\\
 TPR     & 0.00847 & 0.1871 & 0.0813 & 0.1111 & 0.3488 & 0.1390\\
 F1      & 0.01351 & 0.3053 & 0.1227 & 0.1633 & 0.5042 & 0.2138\\
			\hline
			
		\end{tabular}
		\caption{BLAST vs BLAST+CLIQUE (VFFGDSLSDN)}
		\label{table:table3}		
	\end{center}

\end{table}


\FloatBarrier

\bibliographystyle{vancouver}
\bibliography{Clique.bib}

\end{document}